\documentclass[
  twoside,
  twocolumn,
  aps,
  10pt,
  prb,
  preprintnumbers,
  floatfix,
  showpacs,
  citeautoscript,
]{revtex4-1}

\usepackage{amsmath}
\usepackage{mathtools}
\usepackage{graphicx}
\graphicspath{{./figs/}}
\usepackage{enumitem}
\usepackage{tabularx}
\usepackage{units}
\usepackage{glossaries}
\setacronymstyle{long-short}
\usepackage{hyperref}

\newacronym{ardr}{ARDR}{automatic relevance detection regression}
\newacronym{ase}{\textsc{ASE}}{atomic simulation environment}
\newacronym{dft}{DFT}{density functional theory}
\newacronym{dof}{DOF}{degrees of freedom}
\newacronym{fc}{FC}{force constant}
\newacronym{fcc}{FCC}{face-centered cubic}
\newacronym{fcp}{FCP}{force constant potential}
\newacronym{lasso}{LASSO}{least absolute shrinkage and selection operator}
\newacronym{mae}{MAE}{mean absolute error}
\newacronym{mc}{MC}{Monte Carlo}
\newacronym{md}{MD}{molecular dynamics}
\newacronym{msd}{MSD}{mean-square displacement}
\newacronym{pes}{PES}{potential energy surface}
\newacronym{rmse}{RMSE}{root-mean-square error}
\newacronym{tmd}{TMD}{transition metal di\-chal\-co\-ge\-ni\-de}

\newcommand{\hiphive}{\textsc{hiphive}}
\newcommand{\shengbte}{\textsc{shengBTE}}
\newcommand{\spglib}{\textsc{spglib}}
\newcommand{\phonopy}{\textsc{phonopy}}
\newcommand{\phonothreepy}{\textsc{phono3py}}
\newcommand{\alamode}{\textsc{ALA\-MO\-DE}}
\newcommand{\almabte}{\textsc{almaBTE}}
\newcommand{\aflow}{\textsc{aapl-aflow}}
\newcommand{\sklearn}{\textsc{scikit-learn}}
\newcommand{\tdep}{\textsc{TDEP}}

\newcommand{\sect}[1]{Sect.~\ref{#1}}
\newcommand{\fig}[1]{Fig.~\ref{#1}}

\newcommand{\eq}[1]{Eq.~\eqref{#1}}
\newcommand{\Eq}[1]{Equation~\eqref{#1}}
\newcommand{\tab}[1]{Table~\ref{#1}}

\renewcommand{\vec}[1]{\ensuremath\boldsymbol{#1}}
\renewcommand{\epsilon}[0]{\varepsilon}
\DeclarePairedDelimiter{\norm}{\lVert}{\rVert}

\setlist[itemize]{leftmargin=*, itemsep=0mm}
\setlist[enumerate]{leftmargin=*, itemsep=0mm}

\newcommand{\phys}{
  Chalmers University of Technology,
  Department of Physics,
  Gothenburg, Sweden
}

\begin{document}

\title{
  The \hiphive{} package for the extraction of \texorpdfstring{\\}{}
  high-order force constants by machine learning
}

\author{Fredrik Eriksson}
\author{Erik Fransson}
\author{Paul Erhart}
\email{erhart@chalmers.se}
\affiliation{\phys}

\begin{abstract}
The efficient extraction of force constants (FCs) is crucial for the analysis of many thermodynamic materials properties.
Approaches based on the systematic enumeration of finite differences scale poorly with system size and can rarely extend beyond third order when input data is obtained from first-principles calculations.
Methods based on parameter fitting in the spirit of interatomic potentials, on the other hand, can extract FC parameters from semi-random configurations of high information density and advanced regularized regression methods can recover physical solutions from a limited amount of data.
Here, we present the \hiphive{} Python package, that enables the construction of force constant models up to arbitrary order.
\hiphive{} exploits crystal symmetries to reduce the number of free parameters and then employs advanced machine learning algorithms to extract the force constants.
Depending on the problem at hand both over and underdetermined systems are handled efficiently.
The FCs can be subsequently analyzed directly and or be used to carry out e.g., molecular dynamics simulations.
The utility of this approach is demonstrated via several examples including ideal and defective monolayers of MoS$_2$ as well as bulk nickel.
\end{abstract}

\maketitle

\section{Introduction}

The vibrational properties of solids are pivotal for a large number of physical phenomena, including phase stability and thermal conduction.
In crystalline solids the vibrational motion of the atoms is periodic and commonly described using phonons -- quasi-particles that represent collective excitations of the lattice.
At the first level of approximation, phonons can be obtained within the harmonic limit, which implies non-interacting quasi-particles with infinite lifetimes.
This approach, along with its so-called quasi-harmonic extension, already provides a wealth of information.
There are, however, countless examples where anharmonic effects are crucial and must be accounted for in order to capture the correct physical behavior of a system.
Notable examples include the lattice contribution to the thermal conductivity or the vibrational stabilization of metastable phases, e.g., the body-centered cubic phase of the group IV elements (Ti, Zr, Hf) or the cubic phase of zirconia \cite{ThoVan13}.
In more general terms, phonon-phonon coupling must be taken into account, which leads to finite phonon lifetimes and temperature-dependent frequencies.

Formally, the analysis of vibrational material properties requires a set of \glspl{fc}, which allows the computation of atomic forces solely based on the displacements of atoms from their reference positions.
The harmonic approximation requires only knowledge of the second-order \glspl{fc}, which can be readily extracted using software packages such as \phonopy{} \cite{TogTan15}.
Third-order \glspl{fc}, which are required e.g., for computing the thermal conductivity to the lowest permissible order of permutation theory, can be constructed using, e.g., \phonothreepy{} \cite{TogChaTan15}, \shengbte{} \cite{LiCarKat14}, \almabte{} \cite{CarVerKat17}, and \aflow{} \cite{PlaNatUsa17}.
These packages employ finite differences and a systematic enumeration of atomic displacements, while reference forces are usually obtained using \gls{dft} calculations.\footnote{
    It is also possible to obtain \glspl{fc} from density functional perturbation theory calculations as implemented, for example, in the \textsc{abinit} \cite{GonJolAbr16} and \textsc{quantum-espresso} packages \cite{GiaAndBru17}.
    The latter approach is, however, commonly limited to second and third-order \glspl{fc} as the computation of higher-order terms, by power of the $2n+1$ theorem (see e.g., Ref.~\onlinecite{GonVig89}), would require knowledge of second and higher-order derivatives of the wave function, which are not commonly available in these codes.
}

\Glspl{fc} beyond third-order are required to describe, e.g., me\-ta\-stable systems or the temperature dependence of phonon modes.
The number of \glspl{dof} in higher-order \glspl{fc} increases exponentially with the interaction range and the latter are hence increasingly difficult to extract by enumeration schemes.
Alternatively, one can employ regression schemes \cite{EsfSto08, And12}, as implemented in the \alamode{} \cite{TadGohTsu14} and \tdep{} codes \cite{HelAbr13}.
They employ use linear lea\-st-square fitting and thus require the number of input forces to exceed the number of parameters, i.e. they solve an \emph{overdetermined} problem.
More recently, techniques based on compressive sensing \cite{CanWak08} have been proposed \cite{ZhoNieXia14, TadTsu15, ZhoNieXia18, ZhoSadAbe18} that can also efficiently solve \emph{underdetermined} systems.

Here, we introduce the \hiphive{} Python package, which allows one to efficiently obtain high-order \glspl{fc} both in large systems and systems with low-symmetry.
\hiphive{} can take advantage of various powerful machine learning algorithms for \gls{fc} extraction via \sklearn{} \cite{PedVarGra11} and it can be readily interfaced with a large number of electronic structure codes via the \gls{ase} package \cite{LarMorBlo17}.
This enables a flexible workflow with easy access to various advanced optimization techniques some of which are designed to find sparse solutions \cite{NelHarZho13, ZhoNieXia14} that reflect the short-range nature of the \glspl{fc}.
If the input configurations are constructed sensibly this approach requires a \emph{much} smaller number of input configurations and thus considerably reduces the computational effort, the overwhelming part of which is usually associated with \gls{dft} calculations.
This approach becomes genuinely advantageous for obtaining second-order \glspl{fc} in large and/or low symmetry systems (defects, interfaces, surfaces, large unit cells etc) and high order \glspl{fc}, for which a strict enumeration scheme quickly leads to a dramatic increase in the number of force calculations.

\Glspl{fc} can be post-processed in a number of ways including analysis via \phonopy{} and \phonothreepy{} as well as \gls{md} simulations via \gls{ase}.
An extensive user guide is available online \cite{homepage}, which includes a basic tutorial as well as a number of advanced examples.
The package is maintained under an open source license on \textsc{gitlab} and can be installed from the \textsc{PyPi} index.

The remainder of this paper is organized as follows.
The next section provides a concise summary of \gls{fc} expansions, which sets up a description of algorithmic and methodological aspects in \sect{sect:methodology} as well as the \hiphive{} workflow in \sect{sect:work-flow}.
Possible applications are finally illustrated by several examples in \sect{sect:examples}.

\section{Force constants}
\label{sect:force-constants}

This section provides a brief introduction to \gls{fc} expansions.
The similarities between a \gls{fc} and a cluster expansion are described and the effect of crystal symmetries is demonstrated.
Finally, the constraints due to translational and rotational symmetry are summarized.

\subsection{Basics}

The potential energy $V$ of a solid can be represented by a Taylor expansion of the \gls{pes} in ionic displacements $\vec{u}=\vec{R}-\vec{R}_0$ away from the equilibrium positions $\vec{R}_0$
\begin{align*}
  V &=
  V_0
  + \Phi_{i}^{\alpha} u_i^\alpha
  + \frac{1}{2}\Phi_{ij}^{\alpha\beta} u_i^\alpha u_j^\beta
  + \frac{1}{3!}\Phi_{ijk}^{\alpha\beta\gamma} u_i^\alpha u_j^\beta u_k^\gamma
  + \dots,
\end{align*}
where the Einstein summation convention applies and
\begin{align*}
  \Phi_{i}^{\alpha} = \frac{\partial V}{\partial u_i^\alpha}
  ,\quad
  \Phi_{ij}^{\alpha\beta} = \frac{\partial^2V}{\partial u_i^\alpha \partial u_j^\beta}
  \quad
  \text{etc.}
\end{align*}
$\Phi_{i}$, $\Phi_{ij}$, \ldots{} are the \glspl{fc} corresponding to increasing orders of the expansion.
Latin indices $i$ indicate the atomic labels, where the summation is over an infinite crystal lattice, while Greek indices $\alpha$ run over the Cartesian coordinates $x, y, z$.
$V_0$ is a constant term, which is commonly ignored when dealing with lattice dynamics.
The first order \gls{fc} is also often dropped since the expansion in displacements can be made around an equilibrium lattice configuration with vanishing forces.
These two terms are important for some applications but here are considered zero.
Truncating the potential after the second-order term results in the conventional harmonic phonon theory, which is analytically solvable and widely used \cite{BorHua54, Wal98}.

The forces can be written in terms of the \glspl{fc} as
\begin{align}
  F_i^\alpha =
  - \Phi_{ij}^{\alpha\beta}  u_j^\beta
  - \frac{1}{2}\Phi_{ijk}^{\alpha\beta\gamma}  u_j^\beta u_k^\gamma
  - \dots
  \label{eq:force}
\end{align}
Crucially, this expression that takes the functional form of an interatomic many-body potential is linear in the \glspl{fc}, which will become relevant in \sect{sect:methodology}.

The number of \glspl{dof} in the \gls{fc} expansion scales exponentially with $\mathcal{O}(N^n)$, where $N$ is the number of atomic sites of the supercell and $n$ is the order after which the expansion is truncated.
The number of \emph{independent} parameters is, however, much smaller due to the symmetries of the underlying lattice as well as constraints due to the conservation of linear and angular momentum \cite{EsfSto08}.
The number of \emph{numerically relevant} parameters is smaller still due to the decay of the \glspl{fc} with interaction distance, order, and finite many-body interactions.

\subsection{Clusters}

\begin{figure}
    \centering
    \includegraphics[width=0.98\columnwidth]{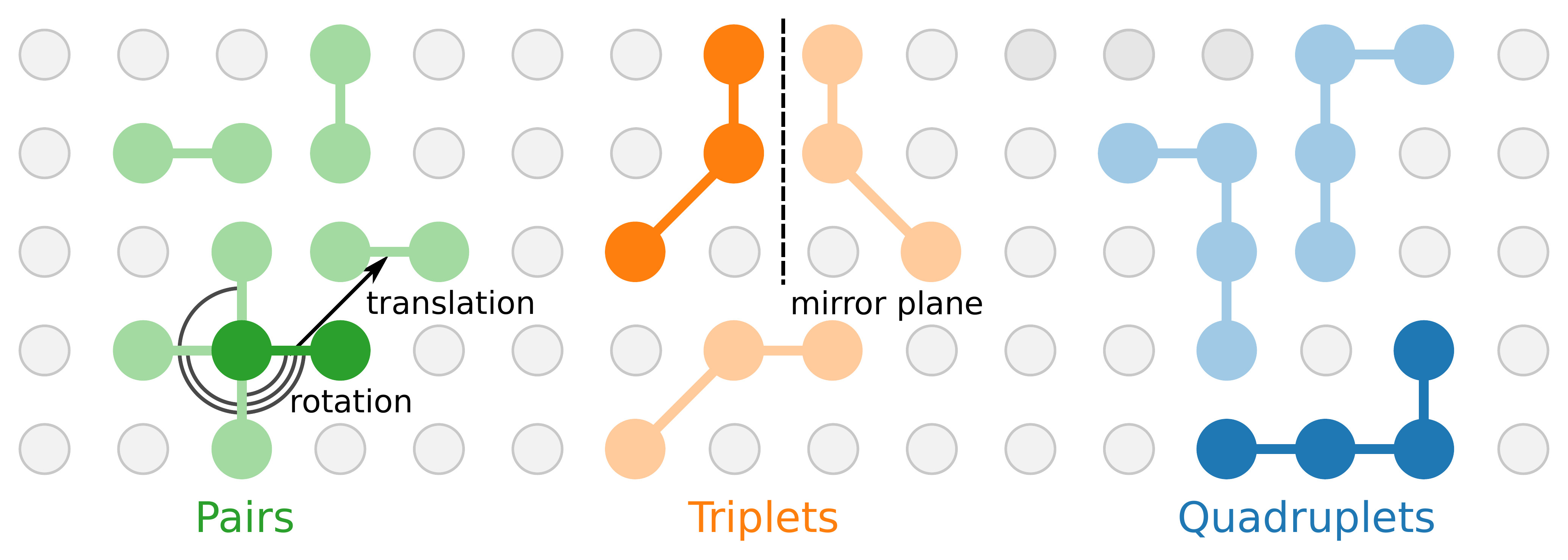}
    \caption{
      Examples for one-body (singlet; red), two-body (pair; green), and three-body (triplet; blue) clusters.
      Dark colors indicate representative cluster whereas lighter colors represent symmetry equivalent clusters that belong to the associated orbit.
      While the two-body clusters shown here are related by a simple fourfold rotation, the three-body clusters can be mapped onto each by application of a fourfold rotation combined with a mirror operation.
    }
    \label{fig:clusters}
\end{figure}

To represent the \glspl{fc} and their inherent symmetries it is convenient to consider clusters of sites $(ij\ldots)$, each of which can be assigned a size, taken for example as the largest distance between any two sites in the cluster.
The \glspl{fc} correspond to interactions between the sites forming a cluster; for example, the term $\Phi_{ijkk}$ describes a fourth-order interaction in the three-body cluster $(ijk)$.
Clusters can be categorized based on the number of sites they comprise and accordingly there are one-body (singlet), two-body (pair), and $n$-body (many-body) clusters (\fig{fig:clusters}).
In general a cluster of order $n$ is a multiset of order (or cardinality) $n$ consisting of up to $n$ different atoms.
As a result of the locality of the interactions, one commonly truncates the \gls{fc} expansion both in size, order, and cluster size.
It is thus possible to create for example a non-central, short ranged, anharmonic pair potential or a long ranged harmonic pair potential with short ranged anharmonic many-body corrections.

\subsection{Symmetries}

The \glspl{fc} obey a number of symmetries, the most fundamental of which stems from the requirement that the differentiation of the total energy and thus $\Phi$ must be invariant under a simultaneous permutation $P$ of atomic and Cartesian indices
\begin{align}
  \label{eq:diff-sym}
  \Phi_{ij\dots}^{\alpha\beta\dots} &= \Phi_{P(ij\dots)}^{P(\alpha\beta\dots)}
\end{align}

The \glspl{fc} must further comply with the symmetry of the underlying lattice as expressed by the associated spacegroup.
Each symmetry operation $S$ consists of a (possibly improper) rotation $R$ and a translation $T$.
If application of $S$ maps a cluster $(ij\dots)$ onto another cluster $(i'j'\dots)$ the \glspl{fc} corresponding to these two clusters are related to each other according to
\begin{align}
  \label{eq:orbit-sym}
  \Phi_{ij\dots}^{\alpha\beta\dots}
  &=
  \Phi_{i'j'\dots}^{\alpha'\beta'\dots}R^{\alpha'\alpha}R^{\beta'\beta}\dots
\end{align}
Clusters, which can be transformed into each other as a result of the application of such a symmetry operation, are said to belong to the same \emph{orbit} (\fig{fig:clusters}) and
the associated \gls{fc} parameters are related by the respective symmetry operation.
The lattice symmetry can thus be used to reduce the number of free parameters by grouping clusters in orbits.

Furthermore, a symmetry operation that maps a cluster onto itself implies a reduction in the number of internal \glspl{dof} in the \glspl{fc},
\begin{align}
  \label{eq:self-sym}
  \Phi_{ij\dots}^{\alpha\beta\dots}
  &=
  \Phi_{ij\dots}^{P^{-1}(\alpha'\beta'\dots)}R^{\alpha'\alpha}R^{\beta'\beta}\dots,
\end{align}
where $P$ can be a permutation, e.g., a mapping of the indices $(ij)$ onto $(ji)$.
\sect{sect:methodology} addresses how these constraints are imposed in practice.

\subsection{Constraints}

The conservation of linear momentum constrains the \glspl{fc} further leading to a set of translational (acoustic)) sum rules.
\begin{align}
    \sum_i\Phi_{ij\dots}^{\alpha\beta\dots} &= 0,
    \label{eq:translational-sym}
\end{align}
which apply for all Cartesian indices independently \cite{BorHua54}.

Rotational invariance is general harder to enforce than translational invariance.
This is partly due to subtle difficulties that arise when combining periodic boundary conditions with a rotationally invariant expansion around equilibrium \cite{CarLiLin16}.
Many authors have described and proposed rotational sum rules \cite{GazWal66, SarSen77, Wal98, WanZhaWu07, ElcEtx11}.
In general it is recognized that the conservation of angular momentum leads to a set of rotational sum rules that relate force constants of order $n$ and $n+1$ as \cite{EsfSto08, Wal98}
\begin{equation}
  \label{eq:rotational-sym}
\begin{aligned}
  \Phi^{\alpha'\dots\alpha_n}_{i_1\dots i_n}\omega_{\gamma}^{\alpha'\alpha_1} + \dots
  \Phi^{\alpha_1\dots\alpha'}_{i_1\dots i_n}\omega_{\gamma}^{\alpha'\alpha_n} \\
  \qquad + \Phi^{\alpha_1\dots\alpha_{n+1}}_{i_1\dots i_{n+1}}\omega_{\gamma}^{\alpha_{n+1}\alpha'}r_{i_{n+1}}^{\alpha'} = 0,
\end{aligned}
\end{equation}
where $\omega$ is a generator of infinitesimal rotations with the same representation as the Levi-Civita symbol and $r_i^\alpha$ denotes the position vector of atom $i$.

To simplify things, we only consider the two simplest sum rules derived by Born and Huang \cite{BorHua54} and apply them to second order \glspl{fc}.
The Born-Huang sum rule reads
\begin{align}
\label{eq:Born-Huang}
    \Phi_{ij}^{\alpha\beta}r_j^\gamma = \Phi_{ij}^{\alpha\gamma}r_j^\beta,
\end{align}
which is a truncation of \eq{eq:rotational-sym}, whereas the Huang invariance imposes
\begin{align}
\label{eq:Huang-invariances}
    \sum_{ij}\Phi_{ij}^{\alpha\beta}r_{ij}^\gamma r_{ij}^\delta
    =
    \sum_{ij}\Phi_{ij}^{\gamma\delta}r_{ij}^\alpha r_{ij}^\beta,
\end{align}
where $r_{ij}$ is the distance vector between the equilibrium positions of atoms $i$ and $j$.

These two constraints do not suffice to render the expansion fully rotationally invariant and thus \gls{md} simulations can potentially behave unphysical.
As shown below, they do, however, enforce the correct dispersion relation near the $\Gamma$-point, which is crucial for e.g., two-dimensional materials \cite{CarLiLin16}.

\begin{figure}
    \centering
    \includegraphics[scale=1.0]{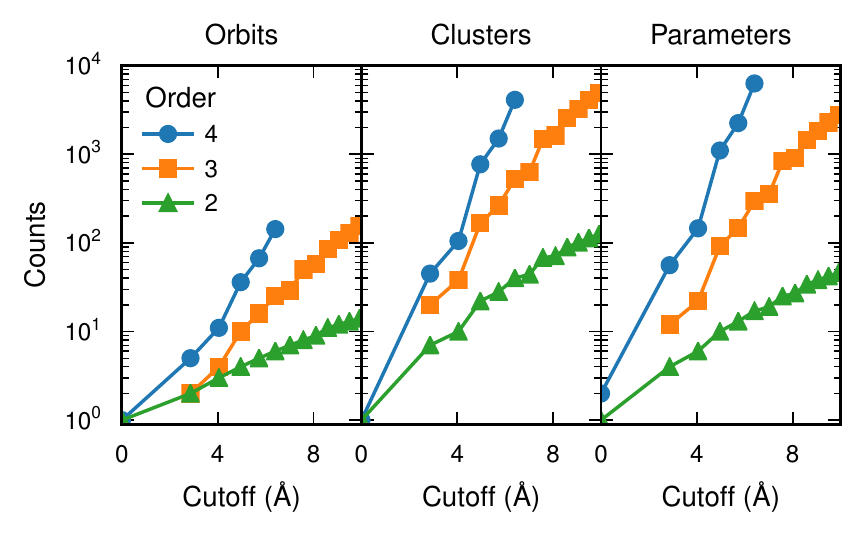}
    \caption{
      (a) Number of unique types of clusters,
      (b) total number of clusters per primitive cell, and
      (c) number of unknown parameters in \gls{fcc} aluminum as a function of the cutoff radius imposed during construction of the cluster space.
    }
    \label{fig:cluster-analysis}
\end{figure}

Even with the application of symmetry and sum rules the number of orbits increases rapidly with cutoff and order [\fig{fig:cluster-analysis}].
Since each orbit is associated with several parameters, the total number of parameters increases even more quickly.
As a result for systems with low symmetry, large unit cells, and high orders the number of model \glspl{dof} can easily outnumber the number available reference forces leading to underdetermined problem as discussed below.

\section{Methodology}
\label{sect:methodology}

This section will start with a concise overview of the computational methods used when implementing the framework described above.
In the subsequent subsections each step is described in more detail.

The \emph{cluster space} represents a fundamental element of the algorithm implemented in \hiphive.
It comprises information concerning
\begin{itemize}
  \item the clusters that an atom in the primitive cell can be part of given the cutoff,
  \item the organization of clusters into orbits,
  \item the \gls{fc} for each orbit (relative to a representative cluster of that orbit),
  \item the free parameters allowed by symmetry in each \gls{fc}, and
  \item the constraints imposed on the free parameters by translational and rotational invariance.
\end{itemize}
The cluster space thus contains all the information needed to construct the template \glspl{fc} for any supercell of the same structure.

The information is generated by the following procedure, which requires user input in the form of a structure (defined by cell metric, basis, and chemical species) as well as a set of cutoff radii per order.
\begin{enumerate}
\item
  Find all applicable symmetry operations for the input structure using, e.g., \spglib{} \cite{spglib}.
\item
  Enumerate all possible sites in periodic images of the primitive cell located at the origin (zero-cell) that reside within the specified cutoff relative to any atom in the center cell
\item
  Construct all possible clusters, which contain at least one site in the zero-cell and are consistent with the cutoff.
\item
  Apply symmetry operations and establish the symmetry relations between clusters; group the clusters into orbits.
\item
  Use the symmetry operations that map the representative cluster of an orbit onto itself to reduce the \glspl{fc} into reduced components that obey the symmetries.
\item
  Finally, construct the system of equations describing the sum rules and find the solution space satisfying the constraints.
\end{enumerate}
The cluster space can then be used together with a supercell with displaced atoms to create the so called sensing matrix, which relates the free parameters to the resulting forces.

\subsection{Clusters and orbits}

Finding the possible clusters in a structure can be done in several ways.
Keep in mind that even though a lower order cluster (e.g., 0-0-3-7) might be invariant under a symmetry, a higher-order cluster involving the same atoms (e.g. 0-3-3-7-7) need not be invariant under the same symmetry.
One simple way to generate clusters is to generate all possible unique multisets of increasing cardinality with elements drawn from the atom enumeration of the periodic images.
To decide if the cluster (multiset) should be kept all atoms must be within the cutoff distance from each other and it must contain at least one atom in the zero-cell.

The symmetry operations can now be used to categorize the clusters into orbits, each of which contains a representative cluster.
All other clusters in the orbit are related to the representative (prototype) cluster by a symmetry operation together with a permutation.
If during the categorization process a symmetry maps a cluster onto a permutation of itself it can be used to reduce the number of free components in the \gls{fc} as described below.

\subsection{Cluster symmetries and invariant bases}

If a representative cluster is symmetric under a set of crystal symmetries $\{S_i\}$ the corresponding representative \gls{fc} must obey the same symmetries.
Using multi-index notation $\cdot^\Lambda=\{\cdot_{ij}^{\alpha\beta}, \cdot_{ijk}^{\alpha\beta\gamma}, \ldots\}$, the problem of finding valid solutions can be transformed into an eigenvalue problem by flattening the tensors
\begin{align}
    \Phi^\Lambda = \Phi^{\Lambda'} \widetilde{R}^{P(\Lambda')\Lambda}.
    \label{eq:eigen-sym}
\end{align}
The eigenvectors, after back-transformation, are then solutions $\varphi$ to the original equation \eqref{eq:self-sym}, which form an invariant basis with respect to the symmetries of the cluster.
A \gls{fc} associated with the respective cluster that fulfills the symmetry can then be constructed as a linear combination of these basis functions
\begin{align}
    \Phi = \sum_p a_p \varphi_p,
    \label{eq:fc-basis}
\end{align}
where $a_p$ represents the parameters that must be found by optimization (see below) and cannot be determined by symmetry alone and the basis elements $\varphi$ we denote as eigentensors.
\Eq{eq:eigen-sym} can in principle be solved by any suitable algorithm.
It is, however, preferable to work in scaled coordinates since then the rotation matrices are integer matrices allowing one to use an optimized algorithm.
This has some additional advantages including exact precision and conservation of sparsity, i.e., a $3\times3$ matrix with 9 unknowns can be decomposed into 9 dense $3\times3$ matrices or 9 sparse $3\times3$ matrices with only one element per matrix.

\subsection{Translational sum rules}

While the parametrization obtained after enforcing the crystal symmetries could in principle be used as the final parametrization of the model, the resulting \glspl{fc} are not guaranteed to be translational or rotational invariant.
The constraints due to sum rules can be enforced by projecting the parameters onto a subspace of the parameter space (i.e. the nullspace of \eq{eq:translational-sym}) that fulfills the sum rule.
As noted below, this method is used in \hiphive{} to enforce the rotational sum rules.

Sum rules can also be applied via re-parametrization by only considering linear combinations of parameter vectors, which span the aforementioned nullspace of the sum rule.
\hiphive{} adopts this approach to enforce the translational sum rules.
The corresponding constraint matrix is constructed as described below.

Let $\Theta$ denote an orbit and $T_{\Theta c}$ the combined rotation and permutation operation that maps the representative cluster $c_\Theta$ onto a cluster $c$ belonging to the same orbit.
Then together with \eq{eq:fc-basis} the sum rule \eq{eq:translational-sym} can be written as
\begin{align}
\begin{matrix}
    \sum_i \sum_\Theta \sum_p a_{p\Theta} \varphi_{p\Theta}^{\alpha'\beta'\dots}
    T^{\alpha'\beta'\dots\alpha\beta\dots}_{\Theta (ij\dots)} = 0 \quad\forall j\dots\alpha\beta\dots\\
    =\boldsymbol{C}^\text{trans} \boldsymbol{a} = \boldsymbol{0}
\end{matrix}
\end{align}
where $\boldsymbol{C}^\text{trans}$ is the translational constraint matrix.
Interpreting $p\Theta$ as a multi-index the equation above represents a system of linear equations in the parameters $a_{p\Theta}$.
The solutions to this system are referred to as constraint vectors $A_{p\Theta,i}$ and specify how the parameters $a_{p\Theta}$ must be related in order to fulfill the sum rules.
\[
a_{p\Theta} = A_{p\Theta,i} \tilde{a}_i
\]
where $\tilde{a}$ is the new parameters guaranteed to fulfill the translational sum rules.

\subsection{Sensing matrix}

To extract the independent parameters the so called sensing (or fit) matrix $\boldsymbol{M}$ must be constructed for each input structure.
This matrix, which depends on the $3N_\text{at}$-dimensional displacement vector $\boldsymbol{u}$, relates the $3N_\text{at}$-dimensional vector of predicted forces $\boldsymbol{f}$ to the $N_\text{par}$-dimensional parameter vector $\boldsymbol{\widetilde{a}}$,
\begin{equation}
    \label{eq:sensing}
    \boldsymbol{f}(\boldsymbol{u},\boldsymbol{\widetilde{a}}) = \boldsymbol{M}(\boldsymbol{u}) \boldsymbol{\widetilde{a}},
\end{equation}
where $N_\text{at}$ and $N_\text{par}$ denote the number of atoms in the supercell and the number of parameters, respectively.
This form is possible due to \eq{eq:force}, which has the form of an interatomic potential with tunable parameters.
The crucial step here is to recognize that the forces $\boldsymbol{f}$ are linearly related to the \gls{fc} parameters $\boldsymbol{\widetilde{a}}$ for given displacements $\boldsymbol{u}$.
The \glspl{fc} in turn are linearly dependent on the \gls{fc} parameters via the crystal symmetries and the \gls{fc} parameters associated with the orbits, where the parameters are either the true expansion parameters associated with the orbits or the constrained parameters that include the translational sum rules.

\subsection{Rotational sum rules}

In principle the procedure used to enforce the translational sum rules can also be employed to apply the rotational sum rules.
Since the rotational sum rules involve the positions of the lattice sites (which can assume any real value) the algorithm to extract the nullspace must, however, be numerically very robust.
Alternatively, one can project a previously determined parametrization onto the correct subspace while maintaining lattice symmetries and translational invariance, which is the approach adopted here.

The sum rules, e.g.,
\begin{align*}
  \Phi_{ij}^{\alpha\beta}r_j^\gamma - \Phi_{ij}^{\alpha\gamma}r_j^\beta = 0,
\end{align*}
are constructed and flattened to a column in a new constraint matrix.
This is repeated for all parameters $\boldsymbol{\widetilde{a}}$ leading to a $N_\text{prim}\times \widetilde{N}_\text{par}$ matrix where $N_\text{prim}$ is the number of atoms in the primitive cell times $3^4$ and $\widetilde{N}_\text{par}$ is equal to the number of independent parameters after the enforcement (re-parametrization) of the translational sum rules.
The sum rules using this new constraint matrix $\boldsymbol{C}^\text{rot}$ and the  parameters $\boldsymbol{\widetilde{a}}$ is written analogous to the translational sum rules as
\begin{align*}
  \boldsymbol{C}^\text{rot} \boldsymbol{\widetilde{a}} = \boldsymbol{0}.
\end{align*}
Given a solution $\boldsymbol{\widetilde{a}}$, the above is in general not fulfilled.
Let us assume $\boldsymbol{C}^\text{rot} \boldsymbol{\widetilde{a}}=\boldsymbol{d}$, where $\boldsymbol{d}$ is a vector describing how well the sum rule is fulfilled.
Now we want to find a correction $\boldsymbol{\Delta\widetilde{a}}$ to $\boldsymbol{\widetilde{a}}$, which is as small as possible and ensures the sum rules are fulfilled,
\begin{align}
\boldsymbol{C}^\text{rot} (\boldsymbol{\widetilde{a}}+\boldsymbol{\Delta\widetilde{a}})
  = \boldsymbol{0}.
\end{align}
Thus we have a new problem, i.e. to find $\boldsymbol{\Delta \widetilde{a}}$ such that $\norm{\boldsymbol{C}^\text{rot}\boldsymbol{\Delta \widetilde{a}} + \boldsymbol{d}}<\epsilon_1$ while $\norm{\boldsymbol{\Delta \widetilde{a}}}<\epsilon_2$, where $\epsilon_i$ are numerical tolerance parameters.
This problem can be solved efficiently by e.g., ridge regression using the $\ell_2$-norm.

\subsection{Reference structures}
\label{sect:training-structures}

Extracting \glspl{fc} requires a set of structures with reference forces that are most commonly obtained from \gls{dft} calculations.
The compilation of these structures depends on the intended usage.
In the present context, we are usually faced with \gls{fc} extraction in one of the following situations.
\begin{enumerate}
\item
    \glspl{fc} to be used in lattice dynamics theory (phonon frequencies, lifetimes etc and derived properties such as harmonic free energy and thermal conductivity)
\item
    \glspl{fc} to be used as an interatomic potential in e.g., \gls{md} simulations for sampling strongly anharmonic \glspl{pes}
\item
    effective lower-order \glspl{fc} directly fitted to a \gls{md} trajectory
\end{enumerate}

In the last case, structure selection is trivial as the \gls{md} trajectory naturally delivers the structures of interest.
The other two scenarios require, however, more attention.

In case (i) often only second and third order \glspl{fc} are of interest.
For this purpose, we have found ``rattled'' structures to work well, which can be obtained by imposing displacements drawn from a Gaussian distribution.
Displacement amplitudes of $\sim$ 0.01 to \unit[0.05]{\AA} commonly yield accurate force constants, which is comparable to the default displacement amplitude of \unit[0.01]{\AA} used for example by \phonopy{}.

When using the rattle approach we have found that it can become difficult to untangle the force contribution if higher orders are contributing to the forces.
This leads to higher order contributions effectively being included in the fitted force constants thus reducing their accuracy.
Therefore, even if the goal is to extract only second or third-order \glspl{fc} it is beneficial to include terms corresponding to the respective higher order (third or fourth-order) during \glspl{fc} extraction.

In case (ii) it is more difficult to produce good training structures without any prior knowledge of the \gls{pes}.
When constructing reference structures for a fourth, sixth or even higher order model, rather large displacements must be included in training set.
Here, the rattle method with standard deviations of $\gtrsim$ \unit[0.1]{\AA} will fail as it commonly leads to some very short interatomic distances with huge repulsive forces.
This can be overcome using a \gls{mc} approach, which aims to produce large random displacement while preserving interatomic distances.
A randomly selected atom is displaced by a small amount and the new position is accepted with probability
\begin{align*}
    P = \frac{1}{2}\left [ \text{erf}\left( \frac{d_\text{min}-d_\text{threshold}}{d_\text{width}} \right) +1 \right ].
\end{align*}
where $d_\text{min}$ is the new minimum distance to any another atom, $d_\text{threshold}$ is a user defined threshold for interatomic distances and $d_\text{width}$ acts as an effective temperature typically of order $\sim$ 0.1\AA.

There is, however, a more elegant and physically sensible approach.
Using some rough estimate for the second-order \glspl{fc} one can obtain a set of normal modes, which can be subsequently randomly populated with an average energy of $k_BT/2$ to generate a physically sensible displacement pattern \cite{Eck04},
\begin{align*}
    u_i ^\alpha = \sqrt{ \frac{2k_BT}{m_i} } \sum_s \frac{1}{\omega_s} W^\alpha_{is} \sqrt{-\ln Q_s} \cos\left(2\pi U_s\right).
    \label{eq:phonon_rattle}
\end{align*}
where $W$ is the polarization vector of mode $s$, and $Q$ and $U$ are uniform random numbers between zero and one.
This approach can be used to generate displacements corresponding to specific temperatures as shown for example in Ref.~\onlinecite{AbeErhLor13}.
The initial second-order \glspl{fc} can for example be obtained from a fit to a minimal set of rattled structures with a small displacement amplitude.

These two methods allow one to generate sensible training structures with large displacements with minimal knowledge of the reference \gls{pes}.
While we have found these methods to work very well in most cases, it is of course also possible to improve models iteratively.

\section{Workflow}
\label{sect:work-flow}

\begin{figure}
    \centering
    \includegraphics[width=0.98\linewidth]{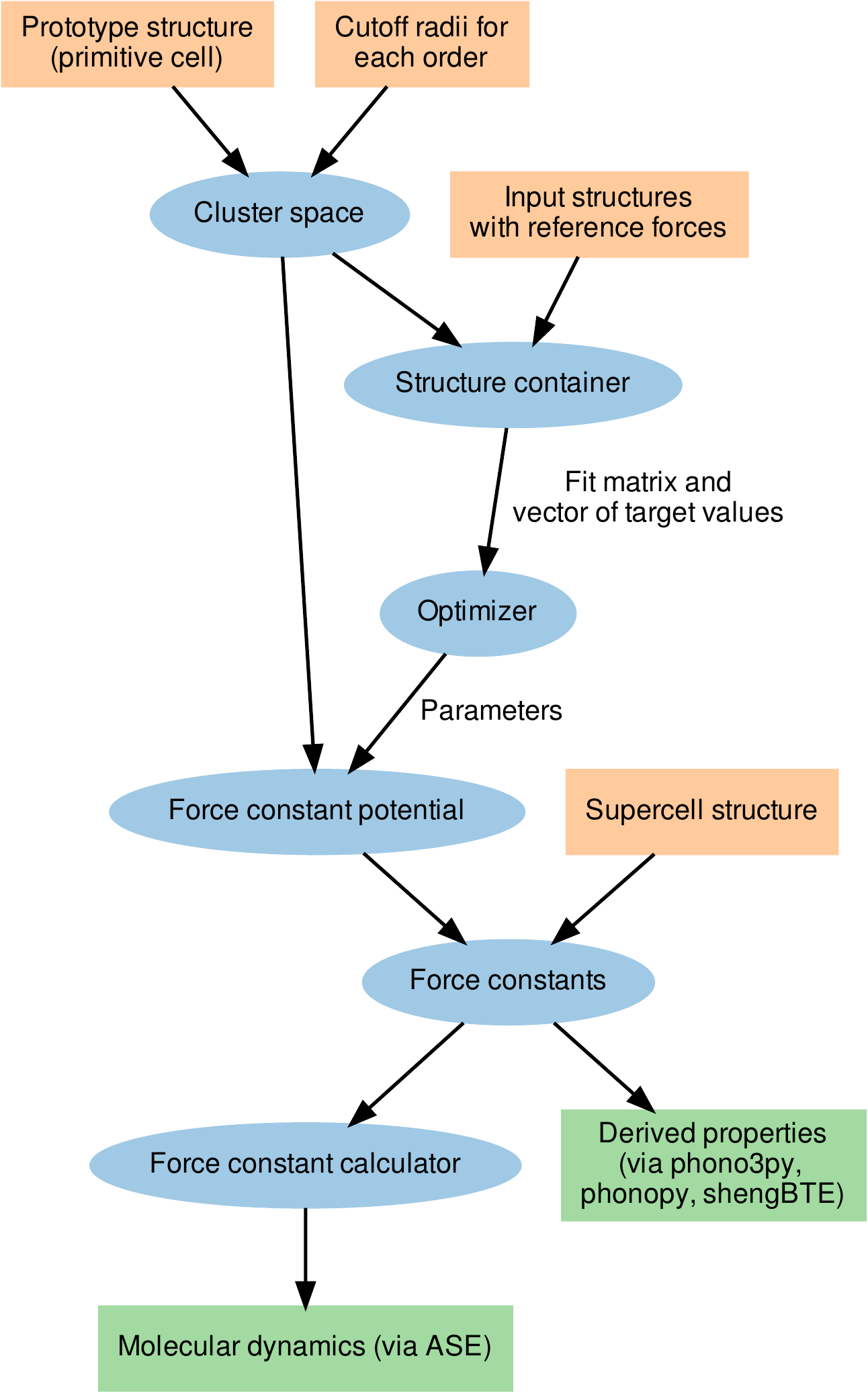}
    \caption{
        Overview of \hiphive{} workflow.
        Square orange nodes represent data supplied by a user.
        Blue ellipses represent \hiphive{} objects.
        Green squares represent output data that can be used directly or processed further using other packages or programs.
    }
    \label{fig:workflow}
\end{figure}

The approach outlined above has been implemented in the Python package \hiphive{}.
The latter also provides functionality for performing related tasks such as fitting and validation, can be interfaced with other Python packages such as e.g., \phonopy{} \cite{phonopy} and \sklearn{} \cite{PedVarGra11}.
The following subsections described the key steps involved in creating a \gls{fcp} (\fig{fig:workflow}).

\subsection{Cluster space}
To build a \gls{fcp}, one must create a cluster space object based on a prototype structure and a set of cutoffs, which specify the maximum interaction range considered for each order.
As described in \sect{sect:methodology}, the cluster space compiles the information needed to completely specify the clusters and eigentensors of any supercell based on a certain primitive cell.
The cluster space is associated with a number of free parameters, which can be extracted by fitting to forces obtained by pseudo-random displacements for a number of supercell structures.

\subsection{Structure container}

In order to fit the parameters, one requires a set of reference forces and displacements for some structures.
The reference structures should span the configuration space of interest and for computational reasons it is desirable to use as few structures as possible (see \sect{sect:training-structures}).

\hiphive{} can handle any combination of structures as long as they share the same equivalent primitive cell.
For convenience, all configurations available for training and validation are compiled into a structure container.
Upon addition of a structure its corresponding sensing matrix is constructed as described in \sect{sect:methodology}.
The structure container then provides functionality for selecting subsets of structures in the form of sensing matrices and target forces suitable for training and validation.

\subsection{Training and validation}

Having constructed cluster space and structure container one can train the associated parameters.
This involves solving a set of linear equations, which can be readily achieved by a number of linear regression techniques.
The equation system can be over or underdetermined and in both cases some form of regularization is useful since the solution is often sparse and/or the data contains noise originating from the input data or the truncation of the cluster space.
The validity of the solution can be assessed by cross validation techniques including.

\begin{figure*}[bt]
  \centering
  \includegraphics[scale=1]{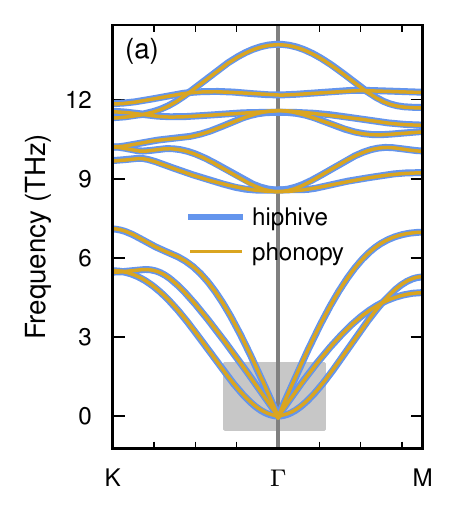}
  \includegraphics[scale=1]{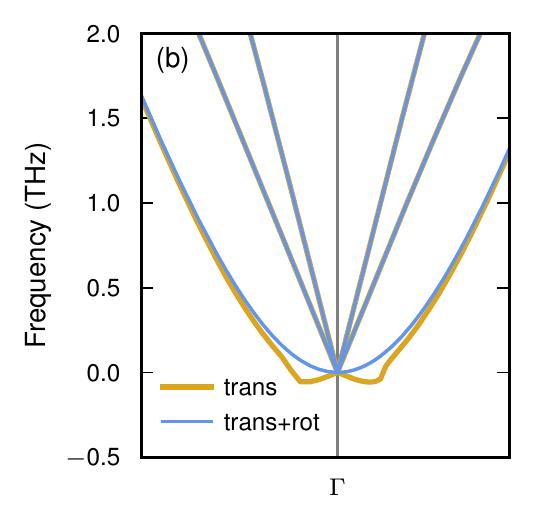}
  \includegraphics[scale=1]{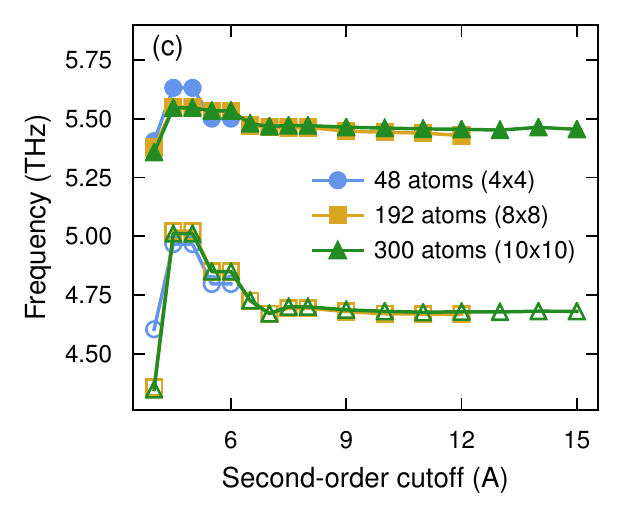}
  \caption{
    Vibrational properties of an ideal monolayer of MoS$_2$.
    (a) Phonon dispersions obtained using \hiphive{} and \phonopy, respectively.
    (b) Effect of rotational invariance conditions on the dispersion in the gray shaded region in (a) near the $\Gamma$-point.
    (c) Convergence of the lowermost modes at K (filled symbols) and M (open symbols) with supercell size and the cutoff for second-order terms.
  }
  \label{fig:ideal-MoS2}
\end{figure*}

\hiphive{} supports a number of popular regularized training and validation techniques via the \sklearn{} machine learning library.
This includes for example methods such as \gls{lasso}, \gls{ardr}, singular-value decomposition, elastic net, Bayesian-ridge regression, and recursive feature elimination.
There is also functionality for generating ensembles of models.
Further training/validation protocols can be readily implemented thanks to the modular structure of the approach.

\subsection{Force constant potential}
Once the free parameters have been obtained the cluster space can be transformed into a \gls{fcp}.
During this step it is also possible to enforce rotational sum rules.
The \gls{fcp} enables calculating the \glspl{fc} in any supercell compatible with the original primitive cell.

The final \glspl{fc} can be analyzed using \phonopy{} to obtain e.g., phonon dispersions or thermodynamic quantities in the (quasi-)harmonic approximation.
It is also possible to compute transport properties using e.g., \phonothreepy{} or \shengbte{} or to carry out \gls{md} simulations via an \gls{ase} calculator.

\section{Applications}
\label{sect:examples}

\subsection{Phonon dispersion of monolayer-MoS\texorpdfstring{$_2$}{2}}

Two-dimensional (2D) materials such as graphene are attracting a lot of interest due their exceptional properties.
As a result of their dimensionality they exhibit a quadratic dispersion of one of the transverse acoustic modes near the $\Gamma$-point \cite{CarLiLin16}.
This is in contrast to (three-dimensional) bulk materials, for which all acoustic branches exhibit a linear dispersion in the center of the Brillouin zone.
It has been shown that in order for this behavior to be captured correctly the \glspl{fc} must fulfill crystal symmetry, translational invariance, as well as \emph{rotational} invariance conditions \cite{CarLiLin16}.
This provides an opportunity for demonstrating the impact of the rotational sum rules on the phonon dispersion.

Here, we consider a monolayer of the \gls{tmd} MoS$_2$, which belongs to space group P$\bar{6}m2$ (International Tables of Crystallography number 187).
Input configurations were generated by imposing random displacements on ideal supercells comprising up to 300 atoms (equivalent to $10\times10\times1$ unit cells).
The average displacement amplitudes for these configurations were approximately \unit[0.008]{\AA} leading to average forces of \unit[170]{meV/\AA} and maximum forces of about \unit[1.1]{eV/\AA}.
Reference forces were obtained from \gls{dft} calculations using the projector augmented wave method \cite{Blo94, KreJou99} as implemented in \textsc{vasp} \cite{KreFur96a} and the vdW-DF-cx method, which combines semi-local exchange with non-local correlation \cite{DioRydSch04, KliBowMic11, BerHyl14, Bjo14}.
The latter has been previously shown to perform very well with regard to the description of both structural and thermal properties of \glspl{tmd} \cite{LinErh16}.
The Brillouin zone was sampled using Monkhorst-Pack $\boldsymbol{k}$-point grids equivalent to a $16\times16\times1$ mesh relative to the primitive cell, except for the 300-atom cells, which were sampled using only the $\Gamma$-point.
The plane-wave energy cutoff was set to \unit[260]{eV}, a finer grid was employed for evaluating the forces, and the reciprocal projection scheme was used throughout.

\Glspl{fcp} were constructed using second-order cutoffs of up to \unit[15]{\AA} and including third-order terms up to a range of \unit[3.0]{\AA}.
The latter procedure was found to stabilize the convergence of the second-order terms.
For the largest supercell (300 atoms) and cutoff radius \unit[15]{\AA} one obtains 319 parameters.
Parameters were optimized by conventional least-squares regression.
For validation the phonon dispersion was also computed using \phonopy{} \cite{TogTan15}.

The phonon dispersion obtained using \hiphive{} with all invariance conditions imposed is virtually indistinguishable from the one generated by \phonopy{} (\fig{fig:ideal-MoS2}a).
Here, the lowermost transverse acoustic branch clearly exhibits a quadratic dispersion.
If the rotational sum rules are deactivated the dispersion is almost unchanged safe for the emergence of a small imaginary pocket in the immediate vicinity of the Brillouin center (\fig{fig:ideal-MoS2}b), and thus does \emph{not} yield a quadratic dispersion.
As noted in Ref.~\onlinecite{CarLiLin16}, this seemingly small error can have a pronounced effect whenever the acoustic modes contribute substantially to a property, as in the case of e.g., the thermal conductivity.

This example also enables us to illustrate the effect of supercell size on the phonon dispersion and thus effectively the range of the \glspl{fc} (\fig{fig:ideal-MoS2}c).
Convergence is achieved at a cutoff of just over \unit[9]{\AA} equivalent to a supercell size of $6\times6\times1$ (108 atoms).

\subsection{Thermal conductivity of monolayer-MoS\texorpdfstring{$_2$}{2}}

\begin{figure*}
  \centering
  \includegraphics[scale=1]{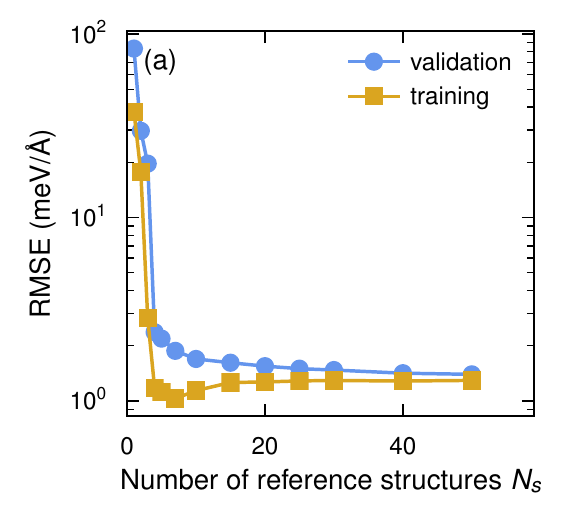}
  \includegraphics[scale=1]{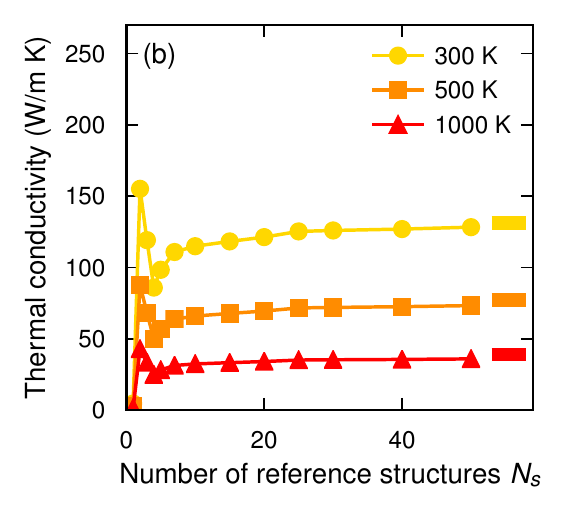}
  \includegraphics[scale=1]{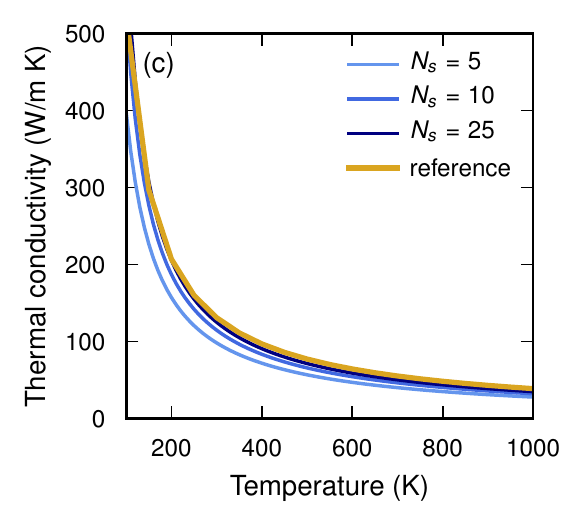}
  \caption{
    Thermal conductivity of an ideal monolayer of MoS$_2$.
    Convergence of (a) \gls{rmse} and (b,c) thermal conductivity with the number of structures in the reference data set.
    The thermal conductivity obtained using the conventional \phonothreepy{} approach is shown by the horizontal bars on the right hand side of panel (b) and the orange line in panel (c).
  }
  \label{fig:kappa-MoS2}
\end{figure*}

Since modeling the thermal conductivity requires both accurate second and third-order \glspl{fc}, it thereby provides a viable test for the extraction of the higher-order \glspl{fc}.
The in-plane thermal conductivity of MoS$_2$ has been studied extensively both experimentally \cite{LiuChoCah14, ZhaSunLi15, JoaQiaGu17} and computationally \cite{LiCarMin13, PenZhaSha16, LinErh16}, and thus constitutes a good test case.

We used 192-atom ($8\times8\times1$) supercells and generated the structures for the reference force calculations were generated by rattling using an average displacement of \unit[0.048]{\AA}.
Reference \gls{dft} calculations were carried out using the same computational parameters as for the ideal monolayer, yielding an average (maximum) force of \unit[993]{mev/\AA} (\unit[4.84]{eV/\AA}).

The cluster space was constructed using cutoffs of \unit[10.0]{\AA}, \unit[6.5]{\AA}, and \unit[3.5]{\AA} for second, third, and fourth-order terms, respectively, yielding 133 orbits, 2209 clusters, and 2004 parameters.
\gls{fcp} parameters were determined using recursive feature elimination with $\ell_2$-minimization.
During training and validation the set of reference forces was split at ratio of 4 to 1 into training and test sets, whereas the final \glspl{fcp} were obtained by fitting against all available data.

The thermal conductivity was evaluated in the framework of phonon Boltzmann transport theory in the relaxation time approximation using the \phonothreepy{} \cite{TogChaTan15} code.
The integration of the Brillouin was carried out using the tetrahedron method and a $40\times40\times1$ $\vec{q}$-point mesh.
For clarity of the analysis, only phonon-phonon scattering was considered as a rate-limiting process.
Reference calculations were carried out including pairs up to a distance of \unit[5.9]{\AA}.

The \gls{rmse} over the validation set quickly drops with the number of structures in the reference set [\fig{fig:kappa-MoS2}], levelling out at a value of about \unit[1.5]{meV/\AA} from about 15 structures onward.
The thermal conductivity converges slightly more slowly as it requires about 20 to 25 configurations to achieve a converged result [\fig{fig:kappa-MoS2}(b)].
Most importantly, the converged values are in very good agreement with the reference data from \phonothreepy{} over the entire temperature range [\fig{fig:kappa-MoS2}(c)], demonstrating the accuracy of the extracted \glspl{fc}.

Crucially this result was achieved at a fraction of the computational cost.
The equivalent \phonothreepy{} calculations require 441 configurations for a cutoff of \unit[5.9]{\AA} (571 configurations using a cutoff of \unit[6.5]{\AA}).
In the case of \shengbte{} \cite{LiCarKat14}, the numbers are somewhat smaller but still comparable in magnitude (324 structures for a cutoff of \unit[5.9]{\AA}).
By contrast, \hiphive{} required only 20 to 25 configurations to reach a converged result.

\subsection{Sulfur vacancy in monolayer-MoS\texorpdfstring{$_2$}{2}}

Defects in general and sulfur vacancies in particular are present in comparably large numbers in monolayers of MoS$_2$ \cite{ZhoZouNaj13} and they can have a very pronounced effect on the thermal conductivity \cite{KatCarDon17}.
To analyze this effect one requires the second-order \glspl{fc} matrix of a defect supercell, which enables one to evaluate the perturbation matrix connecting the ideal and defective systems.
Due to the size of a defect configuration and its low symmetry, computing the \glspl{fc} using the enumeration approach alluded to above commonly requires a large number of individual displacement/force calculations.
Here, we demonstrate that this number can be considerably reduced using \hiphive{} and advanced linear regression techniques.

The sulfur vacancy was described using a 192-atom supercell ($8\times8\times1$ unit cells), which has been found above to yield well-converged \glspl{fc} in the case of the ideal monolayer.
Reference \gls{dft} calculations were carried out using the same computational parameters as for the ideal monolayer.
Structures for reference force calculations were generated by rattling using an average displacement of \unit[0.016]{\AA}, with an average (maximum) force of \unit[330]{meV/\AA} (\unit[1.02]{eV/\AA}).
The cluster space was constructed using cutoffs of \unit[9.0]{\AA} and \unit[3.0]{\AA} for second and third-order terms, respectively, yielding 1,670 orbits, 9,376 clusters, and 13,030 parameters.
\gls{fcp} parameters were determined using \gls{ardr} with a $\lambda$-threshold of $10^4$ and without standardization.
During training and validation the set of reference forces was split at ratio of 4 to 1 into training and test sets, whereas the final \glspl{fcp} were obtained by fitting against all available data.

\begin{figure}
  \centering
  \includegraphics[scale=1]{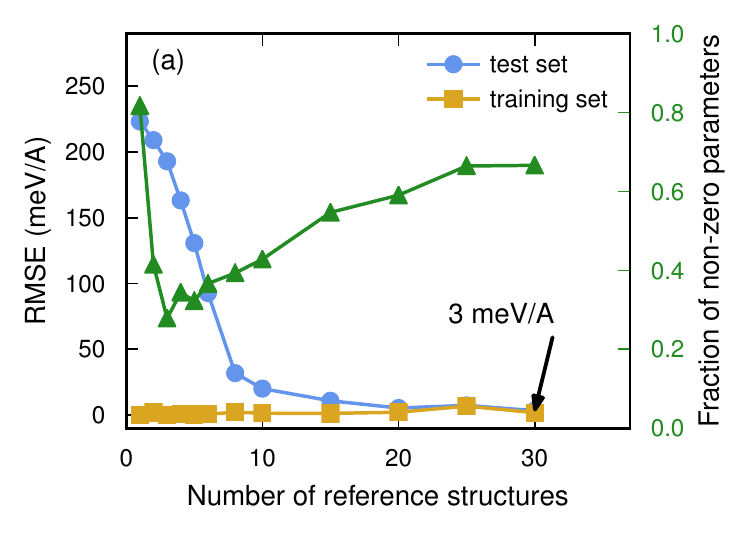}
  \includegraphics[scale=1]{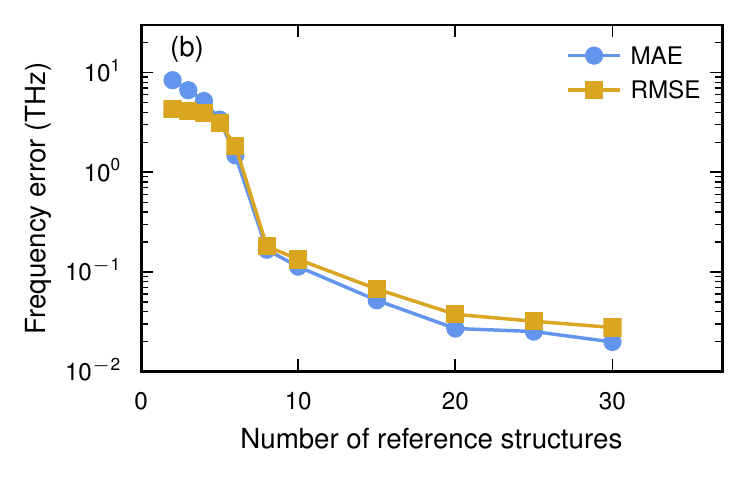}
  \caption{
    Vibrational properties of a sulfur vacancy in a monolayer of MoS$_2$.
    Convergence of the (a) \gls{rmse} and (b) the \gls{mae} and \gls{rmse} with the number of available structures.
  }
  \label{fig:S-vac}
\end{figure}

The \gls{rmse} converges rather quickly with the number of structures (\fig{fig:S-vac}a).
Already for about 25 to 30 structures the \gls{rmse} for the force components in the test set relative to the \gls{dft} reference calculations is only about \unit[3]{meV/\AA}, where the \phonopy{} analysis required 215 individual \gls{dft} calculations.
This convergence behavior also translates to the \gls{mae} and \gls{rmse} over the frequencies (\fig{fig:S-vac}b) as well as the overall phonon dispersion.

The regression produces sparse models as evident from the fraction of non-zero parameters, which remains below 70\%\ (green triangles in \fig{fig:S-vac}a).
This is in line with physical intuition, according to which \glspl{fc} ought to decay rather quickly with interaction distance and order.

\begin{figure}
  \centering
  \includegraphics[scale=1]{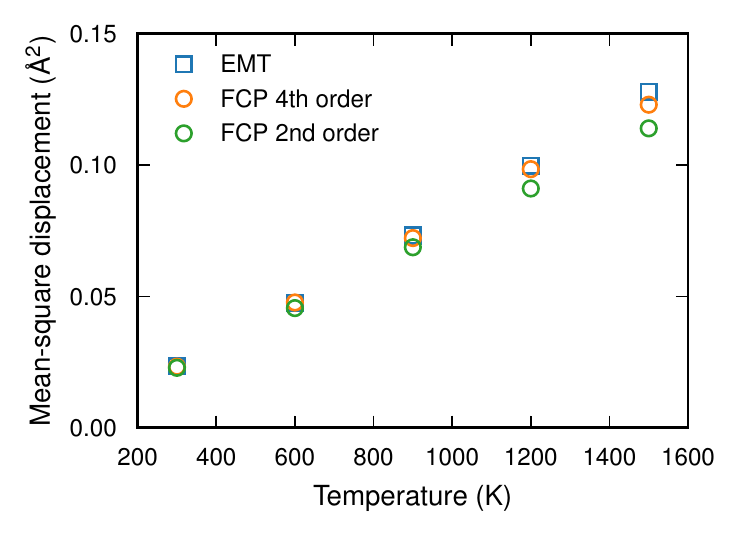}
  \caption{
    Atomic mean-square displacement in \gls{fcc} Ni as a function of temperature as obtained within the harmonic approximation as well as from \gls{md} simulations based on the original potential (EMT) and a fourth-order Hamiltonian.
  }
  \label{fig:msds_nickel}
\end{figure}

\subsection{Molecular dynamics simulations of nickel}

Anharmonic \glspl{fcp} can in principle be sampled using \gls{md} simulations, which provide access to dynamic properties including, e.g., dynamical structure factors, velocity auto-correlation functions, and free energies \cite{dynasor}.
As illustrated in the following, \hiphive{} enables such simulations by providing an \gls{ase} calculator class that merely requires a set of \glspl{fc} as input.

A fourth order \gls{fcp} was created for bulk Ni using cutoff radii of 5.0, 4.0, and \unit[4.0]{\AA} for second, third, and fourth-order terms, respectively, corresponding to 171 clusters in the unit cell.
The associated cluster space contained 20 unique orbits (5, 4, and 11 for second, third, and fourth-order, respectively) with interactions up to the fourth nearest neighbor for the pairs, leading to 119 free parameters.

Training data comprised five structures with 256 atoms ($4\times4\times4$ conventional unit cells) for a total of 3,840 force components.
The structures were obtained by applying displacements randomly drawn from a normal distribution, modified to avoid interatomic distances shorter than \unit[2.3]{\AA}.
The resulting average atomic displacement were about \unit[0.13]{\AA}.
Reference forces were obtained using an effective medium theory model as implemented in \gls{ase} \cite{LarMorBlo17}.
The parameters of the model were trained by standard least-squares fitting as the system is heavily overdetermined (\tab{tab:number_of_orbits}).

\begin{table}
    \centering
    \caption{
        Number of orbits and parameters (in brackets) by order and body.
    }
    \label{tab:number_of_orbits}
    \begin{tabular}{c*{8}c}
        \hline\hline
        & \multicolumn{2}{c}{1-body}
        & \multicolumn{2}{c}{2-body}
        & \multicolumn{2}{c}{3-body}
        & \multicolumn{2}{c}{4-body} \\
        \hline
        2nd order & 1 & (1)  & 4 & (12)  & -- &        & -- &      \\
        3rd order & 0 & (0)  & 2 &  (8)  &  2 & (14)   & -- &      \\
        4th order & 1 & (2)  & 4 & (29)  &  3 & (75)   &  3 & (40) \\
        \hline\hline
    \end{tabular}
\end{table}

The integration of the equations of motion was carried out using functionality provided by \gls{ase} while \hiphive{} was used to provide an interaction model in the form of an \gls{ase} calculator object.
Simulations were carried out using a 864-atom supercell ($6\times6\times6$ conventional unit cells).
The equations of motion were integrated for \unit[50]{ps} using a time step of \unit[5]{fs} at temperatures 300, 600, 900, 1200 and \unit[1500]{K} using a Langevin thermostat.

The atomic mean-square displacements computed using the full fourth-order \gls{fcp} as well as using only the second order \glspl{fc} are practically identical to those obtained using the effective medium theory model at low temperatures (\fig{fig:msds_nickel}).
At higher temperatures where anharmonic effects are more important the second order model yields a systematic error whereas the fourth order model remains very accurate.
Very close to the melting point even the fourth order model starts to deviate from the exact solution, which is due to even higher order terms becoming relevant at these temperatures.

\section{Conclusions and outlook}

Second and higher-order \gls{fc} are fundamental to the description of the thermodynamics of materials.
Here, we have introduced the \hiphive{} package that enables their efficient extraction from first-principles calculations using regression techniques with regularization.
The implementation take advantage of symmetry and sum rules in order to constrain the number of \glspl{dof} and is computationally efficient.
Potential applications have been illustrated by several examples including ideal and defective systems, phonon analysis as well as dynamic simulations.

The package is designed for integration in various workflows including e.g., applications in high-throughput calculations, as it can be readily interfaced with a large number of electronic structure codes via \gls{ase} and machine learning techniques via \sklearn{}.
The flexibility of the interface also enables one to systematically explore the efficacy of different optimization algorithms for the construction of \glspl{fcp} and the compilation of databases of such models for future use.

\section*{Acknowledgments}

This work was funded by the Knut and Alice Wallen\-berg Foundation, the Swedish Research Council as well as the Swedish National Infrastructure for Computing (SNIC).
Computer time allocations by the SNIC at C3SE (Gothenburg), NSC (Link\"oping), and PDC (Stockholm) are gratefully acknowledged.

\end{document}